\documentclass[aps,prl,preprint,superscriptaddress,showpacs]{revtex4-1}
\usepackage{lmodern}%modern Latin fonts: Roman, Sans Serif, Typewriter
\usepackage{amsmath,amssymb,amsthm} %math tool
\usepackage{grffile} %graphic tool <--allows blanks in the file name
\usepackage{sidecap} %side caption tool
\usepackage{enumitem} %list/enumerate item tool
\usepackage{array}
\usepackage{multirow}
\usepackage{ctable} %\specialrule command for tables
\newcommand{\GeV}{$\mathrm{GeV}/{c^{2}}$}
\newcommand{\MeV}{$\mathrm{MeV}/{c^{2}}$}

\begin{document}

% Use the \preprint command to place your local institutional report
% number in the upper righthand corner of the title page in preprint mode.
% Multiple \preprint commands are allowed.
% Use the 'preprintnumbers' class option to override journal defaults
% to display numbers if necessary
%\preprint{preprint submitted to Phys. Rev. Lett.}
\preprint{\vbox{   \hbox{Belle Preprint 2015-13}
                   \hbox{KEK Preprint 2015-22}
}}

%Title of paper
\title{Observation of \boldmath{$B^{0} \rightarrow p\bar{\Lambda} D^{(*)-}$}}

\noaffiliation
\affiliation{University of the Basque Country UPV/EHU, 48080 Bilbao}
\affiliation{Beihang University, Beijing 100191}
\affiliation{University of Bonn, 53115 Bonn}
\affiliation{Budker Institute of Nuclear Physics SB RAS, Novosibirsk 630090}
\affiliation{Faculty of Mathematics and Physics, Charles University, 121 16 Prague}
%%%\affiliation{Chiba University, Chiba 263-8522}
\affiliation{Chonnam National University, Kwangju 660-701}
\affiliation{University of Cincinnati, Cincinnati, Ohio 45221}
\affiliation{Deutsches Elektronen--Synchrotron, 22607 Hamburg}
\affiliation{University of Florida, Gainesville, Florida 32611}
%%%\affiliation{Department of Physics, Fu Jen Catholic University, Taipei 24205}
\affiliation{Justus-Liebig-Universit\"at Gie\ss{}en, 35392 Gie\ss{}en}
\affiliation{Gifu University, Gifu 501-1193}
%%%\affiliation{II. Physikalisches Institut, Georg-August-Universit\"at G\"ottingen, 37073 G\"ottingen}
\affiliation{SOKENDAI (The Graduate University for Advanced Studies), Hayama 240-0193}
\affiliation{Gyeongsang National University, Chinju 660-701}
\affiliation{Hanyang University, Seoul 133-791}
\affiliation{University of Hawaii, Honolulu, Hawaii 96822}
\affiliation{High Energy Accelerator Research Organization (KEK), Tsukuba 305-0801}
%%%\affiliation{Hiroshima Institute of Technology, Hiroshima 731-5193}
\affiliation{IKERBASQUE, Basque Foundation for Science, 48013 Bilbao}
%%%\affiliation{University of Illinois at Urbana-Champaign, Urbana, Illinois 61801}
%%%\affiliation{Indian Institute of Technology Bhubaneswar, Satya Nagar 751007}
\affiliation{Indian Institute of Technology Guwahati, Assam 781039}
\affiliation{Indian Institute of Technology Madras, Chennai 600036}
\affiliation{Indiana University, Bloomington, Indiana 47408}
\affiliation{Institute of High Energy Physics, Chinese Academy of Sciences, Beijing 100049}
\affiliation{Institute of High Energy Physics, Vienna 1050}
%%%\affiliation{Institute for High Energy Physics, Protvino 142281}
%%%\affiliation{Institute of Mathematical Sciences, Chennai 600113}
\affiliation{INFN - Sezione di Torino, 10125 Torino}
\affiliation{Institute for Theoretical and Experimental Physics, Moscow 117218}
\affiliation{J. Stefan Institute, 1000 Ljubljana}
\affiliation{Kanagawa University, Yokohama 221-8686}
\affiliation{Institut f\"ur Experimentelle Kernphysik, Karlsruher Institut f\"ur Technologie, 76131 Karlsruhe}
%%%\affiliation{Kavli Institute for the Physics and Mathematics of the Universe (WPI), University of Tokyo, Kashiwa 277-8583}
\affiliation{Kennesaw State University, Kennesaw GA 30144}
\affiliation{King Abdulaziz City for Science and Technology, Riyadh 11442}
\affiliation{Department of Physics, Faculty of Science, King Abdulaziz University, Jeddah 21589}
\affiliation{Korea Institute of Science and Technology Information, Daejeon 305-806}
\affiliation{Korea University, Seoul 136-713}
%%%\affiliation{Kyoto University, Kyoto 606-8502}
\affiliation{Kyungpook National University, Daegu 702-701}
\affiliation{\'Ecole Polytechnique F\'ed\'erale de Lausanne (EPFL), Lausanne 1015}
\affiliation{Faculty of Mathematics and Physics, University of Ljubljana, 1000 Ljubljana}
\affiliation{Ludwig Maximilians University, 80539 Munich}
\affiliation{Luther College, Decorah, Iowa 52101}
\affiliation{University of Maribor, 2000 Maribor}
\affiliation{Max-Planck-Institut f\"ur Physik, 80805 M\"unchen}
\affiliation{School of Physics, University of Melbourne, Victoria 3010}
\affiliation{Middle East Technical University, 06531 Ankara}
\affiliation{Moscow Physical Engineering Institute, Moscow 115409}
\affiliation{Moscow Institute of Physics and Technology, Moscow Region 141700}
\affiliation{Graduate School of Science, Nagoya University, Nagoya 464-8602}
\affiliation{Kobayashi-Maskawa Institute, Nagoya University, Nagoya 464-8602}
%%%\affiliation{Nara University of Education, Nara 630-8528}
\affiliation{Nara Women's University, Nara 630-8506}
\affiliation{National Central University, Chung-li 32054}
\affiliation{National United University, Miao Li 36003}
\affiliation{Department of Physics, National Taiwan University, Taipei 10617}
\affiliation{H. Niewodniczanski Institute of Nuclear Physics, Krakow 31-342}
%%%\affiliation{Nippon Dental University, Niigata 951-8580}
\affiliation{Niigata University, Niigata 950-2181}
%%%\affiliation{University of Nova Gorica, 5000 Nova Gorica}
\affiliation{Novosibirsk State University, Novosibirsk 630090}
\affiliation{Osaka City University, Osaka 558-8585}
%%%\affiliation{Osaka University, Osaka 565-0871}
\affiliation{Pacific Northwest National Laboratory, Richland, Washington 99352}
%%%\affiliation{Panjab University, Chandigarh 160014}
\affiliation{Peking University, Beijing 100871}
\affiliation{University of Pittsburgh, Pittsburgh, Pennsylvania 15260}
%%%\affiliation{Punjab Agricultural University, Ludhiana 141004}
%%%\affiliation{Research Center for Electron Photon Science, Tohoku University, Sendai 980-8578}
%%%\affiliation{Research Center for Nuclear Physics, Osaka University, Osaka 567-0047}
%%%\affiliation{RIKEN BNL Research Center, Upton, New York 11973}
%%%\affiliation{Saga University, Saga 840-8502}
\affiliation{University of Science and Technology of China, Hefei 230026}
\affiliation{Seoul National University, Seoul 151-742}
%%%\affiliation{Shinshu University, Nagano 390-8621}
\affiliation{Soongsil University, Seoul 156-743}
%%%\affiliation{University of South Carolina, Columbia, South Carolina 29208}
\affiliation{Sungkyunkwan University, Suwon 440-746}
\affiliation{School of Physics, University of Sydney, NSW 2006}
\affiliation{Department of Physics, Faculty of Science, University of Tabuk, Tabuk 71451}
\affiliation{Tata Institute of Fundamental Research, Mumbai 400005}
\affiliation{Excellence Cluster Universe, Technische Universit\"at M\"unchen, 85748 Garching}
\affiliation{Department of Physics, Technische Universit\"at M\"unchen, 85748 Garching}
\affiliation{Toho University, Funabashi 274-8510}
%%%\affiliation{Tohoku Gakuin University, Tagajo 985-8537}
\affiliation{Tohoku University, Sendai 980-8578}
\affiliation{Earthquake Research Institute, University of Tokyo, Tokyo 113-0032}
\affiliation{Department of Physics, University of Tokyo, Tokyo 113-0033}
\affiliation{Tokyo Institute of Technology, Tokyo 152-8550}
\affiliation{Tokyo Metropolitan University, Tokyo 192-0397}
%%%\affiliation{Tokyo University of Agriculture and Technology, Tokyo 184-8588}
\affiliation{University of Torino, 10124 Torino}
%%%\affiliation{Toyama National College of Maritime Technology, Toyama 933-0293}
\affiliation{Utkal University, Bhubaneswar 751004}
\affiliation{CNP, Virginia Polytechnic Institute and State University, Blacksburg, Virginia 24061}
\affiliation{Wayne State University, Detroit, Michigan 48202}
\affiliation{Yamagata University, Yamagata 990-8560}
\affiliation{Yonsei University, Seoul 120-749}

  \author{Y.-Y.~Chang}\email{gixd@hep1.phys.ntu.edu.tw}\affiliation{Department of Physics, National Taiwan University, Taipei 10617} % Taiwan
  \author{M.-Z.~Wang}\email{mwang@hep1.phys.ntu.edu.tw}\affiliation{Department of Physics, National Taiwan University, Taipei 10617} % Taiwan  
  \author{A.~Abdesselam}\affiliation{Department of Physics, Faculty of Science, University of Tabuk, Tabuk 71451} % Tabuk
  \author{I.~Adachi}\affiliation{High Energy Accelerator Research Organization (KEK), Tsukuba 305-0801}\affiliation{SOKENDAI (The Graduate University for Advanced Studies), Hayama 240-0193} % KEK
  \author{K.~Adamczyk}\affiliation{H. Niewodniczanski Institute of Nuclear Physics, Krakow 31-342} % Krakow
  \author{H.~Aihara}\affiliation{Department of Physics, University of Tokyo, Tokyo 113-0033} % Tokyo
  \author{S.~Al~Said}\affiliation{Department of Physics, Faculty of Science, University of Tabuk, Tabuk 71451}\affiliation{Department of Physics, Faculty of Science, King Abdulaziz University, Jeddah 21589} % Tabuk
% \author{K.~Arinstein}\affiliation{Budker Institute of Nuclear Physics SB RAS, Novosibirsk 630090}\affiliation{Novosibirsk State University, Novosibirsk 630090} % BINP
% \author{Y.~Arita}\affiliation{Graduate School of Science, Nagoya University, Nagoya 464-8602} % Nagoya
  \author{D.~M.~Asner}\affiliation{Pacific Northwest National Laboratory, Richland, Washington 99352} % PNNL
% \author{T.~Aso}\affiliation{Toyama National College of Maritime Technology, Toyama 933-0293} % Toyama
  \author{H.~Atmacan}\affiliation{Middle East Technical University, 06531 Ankara} % METU
% \author{V.~Aulchenko}\affiliation{Budker Institute of Nuclear Physics SB RAS, Novosibirsk 630090}\affiliation{Novosibirsk State University, Novosibirsk 630090} % BINP
  \author{T.~Aushev}\affiliation{Moscow Institute of Physics and Technology, Moscow Region 141700}\affiliation{Institute for Theoretical and Experimental Physics, Moscow 117218} % ITEP
% \author{R.~Ayad}\affiliation{Department of Physics, Faculty of Science, University of Tabuk, Tabuk 71451} % Tabuk
% \author{T.~Aziz}\affiliation{Tata Institute of Fundamental Research, Mumbai 400005} % Tata
% ...
% Ljubljana
% \author{N.~Zwahlen}\affiliation{\'Ecole Polytechnique F\'ed\'erale de Lausanne (EPFL), Lausanne 1015} % Lausanne
% \author{O.~Zyukova}\affiliation{Budker Institute of Nuclear Physics SB RAS, Novosibirsk 630090}\affiliation{Novosibirsk State University, Novosibirsk 630090} % BINP
\collaboration{The Belle Collaboration}
 %Belle official author list

%\date{\today}

\begin{abstract}
We report the first observation of the decays $B^0 \rightarrow p\bar{\Lambda} D^{(*)-}$. The data sample of $711$ fb$^{-1}$ used in this analysis corresponds to $772$ million $B\bar{B}$ pairs, collected at the $\Upsilon(4S)$ resonance by the Belle detector at the KEKB asymmetric-energy $e^{+}e^{-}$ collider. We observe $19.8\sigma$ and $10.8\sigma$ excesses of events for the two decay modes and measure the branching fractions of $B^0 \rightarrow p\bar{\Lambda} D^{-}$ and $B^0 \rightarrow p\bar{\Lambda} D^{*-}$ to be $(25.1\pm2.6\pm3.5)\times10^{-6}$ and $(33.6\pm6.3\pm4.4)\times10^{-6}$, respectively, where the first uncertainties are statistical and the second are systematic. These results are not compatible with the predictions based on the generalized factorization approach. In addition, a threshold enhancement in the di-baryon ($p\bar{\Lambda}$) system is observed, consistent with that observed in similar $B$ decays.
\end{abstract}

% insert suggested PACS numbers in braces on next line
% insert suggested keywords - APS authors don't need to do this
%\keywords{Belle; B physics; Baryonic; Threshold effect; Factorization}

%\maketitle must follow title, authors, abstract, \pacs, and \keywords
\maketitle

% body of paper here - Use proper section commands
% References should be done using the \cite, \ref, and \label commands
% Put \label in argument of \section for cross-referencing
%\section{\label{}}
%\section{Introduction}\label{sec_introduction}
In the years since the ARGUS and CLEO collaboration first observed baryonic $B$ decays \cite{Albrecht88,Crawford92}, 
many three-body baryonic $B$ decays ($B\rightarrow\mathrm{\mathbf{B\bar{B}'M}}$) 
have been found \cite{Olive14,Abe02, Aubert06,J.H.Chen08,Wei08}, where $\mathrm{\mathbf{B\bar{B}'}}$ 
denotes a baryon-antibaryon system and $\mathrm{\mathbf M}$ stands for a meson.  Although the general pattern of these decays can be understood as the interplay between the short-distance weak interaction and the long-distance strong interaction \cite{Suzuki07}, theories still have difficulties adjusting for various details such as the angular correlation between the energetic outgoing meson and 
one specific baryon (\textbf{B}) in the di-baryon system \cite{Wang05,Wang07,Wei08,Sanchez12}.

A popular theoretical approach used to investigate the three-body baryonic decays is generalized factorization. 
This method smears the correlation between the weak decay and the fragmentation and allows $B\rightarrow\mathrm{\mathbf{B\bar{B}'M_{c}}}$ decays (with $\mathbf M_{c}$ denoting a charmed meson) to be categorized into three types: current type, where the $\mathbf{B\bar{B}'}$ pair is formed by an external $W$ with other quarks; transition type, where the $W$ is internal and forms $\mathbf{BM_{c}}$; and hybrid (current+transition) type~\cite{C.Chen08}. The
$B^{0} \rightarrow p\bar{\Lambda} D^{(*)-}$~\footnote{Hereafter, the inclusion of the charge-conjugate mode is implied.} decay belongs to the first type whereas its corresponding charged mode, $B^{+} \rightarrow p\bar{\Lambda} \bar{D}^{(*)0}$, 
is of the last type. Using this approach, Ref.~\cite{C.Chen08} predicts the branching fractions 
%\begin{align}\label{BF_thel_B2pLambdaD}
\begin{equation}
\begin{split}
\mathcal{B}(B^{0} \rightarrow p \bar{\Lambda} D^{-})&=(3.4\pm0.2)\times10^{-6},\\ 
\mathcal{B}(B^{0} \rightarrow p \bar{\Lambda} D^{*-})&=(11.9\pm0.5)\times10^{-6},\\ 
\mathcal{B}(B^{+} \rightarrow p\bar{\Lambda} \bar{D}^{0})&=(11.4\pm2.6)\times10^{-6},\\
\mathcal{B}(B^{+} \rightarrow p\bar{\Lambda} \bar{D}^{*0})&=(32.3\pm3.2)\times10^{-6}.
\end{split}
\end{equation}
%\end{align}

There are two salient features of the predicted results. First, the ratios of the branching fractions of the decays into $D^{*}$ to the analogous decays into $D$ are $\approx 3 : 1$. 
Secondly, the branching fraction of the hybrid-type decay is also $\approx 3$ times larger than the corresponding current-type decay.
The measured branching fraction for $B^{+} \rightarrow p\bar{\Lambda} \bar{D}^{0}$ is consistent with the theoretical calculation based on the factorization approach \cite{C.Chen08,Chen11}.

In most $B\rightarrow\mathrm{\mathbf{B\bar{B}'M}}$ decay studies, the final-state di-baryon system is observed to favor a mass near threshold~\cite{Olive14,Bai03,Ablikim05,Alexander10}. While this ``threshold enhancement effect'' is intuitively understood in terms of the factorization approach, such enhancements are not seen in $B^+ \to p \bar{\Lambda} J/\psi$ nor $B^+ \to \Lambda_c^+\Lambda_c^- K^+$ \cite{Xie05, Abe05}. More intriguingly, the factorization approach fails to provide a satisfactory explanation for the $\mathrm{\mathbf M}$--$p$ angular correlations 
in $B^{-} \rightarrow p\bar{p}K^{-}$, $B^{0} \rightarrow p{\bar\Lambda}\pi^-$, 
and $B^{-} \rightarrow p\bar{p}D^{-}$ \cite{Wang05,Wang07,Wei08,Sanchez12}. A striking difference between the non-zero angular asymmetries of $B^{-} \rightarrow p\bar{p}D^{*-}$ and $B^{-} \rightarrow p\bar{p}D^{-}$ was also reported in Ref.\ \cite{Aubert06,C.Chen08}, for which a theoretical explanation was attempted in Ref.\ \cite{Geng06}. A study of pure current-type decays like  $B^{0} \rightarrow p \bar{\Lambda}D^{(*)-}$ is useful to shed more light on the afore mentioned phenomena. In this paper, we report the first observation of $B^{0} \rightarrow p \bar{\Lambda}D^{(*)-}$ decays using data  from the Belle experiment.

%\section{Data sample and signal reconstruction}\label{sec_signal_recon}
The data sample used in this study corresponds to an integrated luminosity of  711 fb$^{-1}$ or $772\times10^{6}$ $B\bar{B}$ pairs produced at the $\Upsilon(4S)$ resonance. The Belle detector is located at the interaction point (IP) of the  KEKB asymmetric-energy $e^{+}$ (3.5 GeV) $e^{-}$ (8 GeV) collider \cite{Kurokawa03,Abe13}. It is a large-solid-angle spectrometer comprising six specialized sub-detectors: the Silicon Vertex Detector (SVD), the 50-layer Central Drift Chamber (CDC), the Aerogel Cherenkov Counter (ACC), the Time-Of-Flight scintillation counter (TOF), the electromagnetic calorimeter, and the $K_L$ and muon detector (KLM). A superconducting solenoid surrounding all but the KLM produces a $1.5$ T magnetic field \cite{Abashian02,Brodzicka12}.

The final-state charged particles, $\pi^{\pm},\ K^{\pm}$ and  % \overset{(-)}{p}$,
\rlap{$\,p$}{${\mathstrut}^{\scriptscriptstyle(-)}$}, 
are selected using the likelihood information from the combined tracking (SVD, CDC) and charged-hadron identification (CDC, ACC, TOF) systems \cite{Nakano02}. The $B^{0} \rightarrow p\bar{\Lambda} D^{(*)-}$ signals are reconstructed through the sub-decays $D^{-}\rightarrow K^{+}\pi^{-}\pi^{-}$, $D^{*-}\rightarrow \bar{D}^{0}\pi^{-}$, $\bar{D}^{0}\rightarrow K^{+}\pi^{-}$, and $\bar{\Lambda}\rightarrow\bar{p}\pi^{+}$. The distance
 of closest approach to the IP by each charged track is required to be less than 3.0 cm along the positron beam ($z$ axis) and 0.3 cm 
in the transverse plane.% Charged tracks are regarded as pions if $\mathcal{L}_{\pi}/(\mathcal{L}_{\pi}+\mathcal{L}_{K})> 0.6$, 
%kaons if $\mathcal{L}_{K}/(\mathcal{L}_{\pi}+\mathcal{L}_{K})> 0.6$, protons if 
%both  $\mathcal{L}_{p}/(\mathcal{L}_{p}+\mathcal{L}_{K})> 0.6$ 
%and $\mathcal{L}_{p}/(\mathcal{L}_{p}+\mathcal{L}_{\pi})> 0.6$, 
%where $\mathcal{L}$ represents the likelihood of the subscript particle determined by the charged hadron identification system. 
The pion and kaon identification efficiencies are in the range of 85--95\% while the probability of misidentifying 
one as the other is 10--20\%, both depending on the momentum. The proton identification efficiency is 90--95\% for the typical momenta in this study, 
and the probability of misidentifying a proton as a pion (kaon) is less than 5\% (10\%). The candidate $\bar{\Lambda}$ is required to have 
a displaced vertex that is consistent with a long-lived particle originating from the IP and an invariant mass between 1.102 and 1.130 \GeV. 
The particle-identification criterion is omitted for the daughter pion in the $\bar{\Lambda}$ reconstruction due to the low background rate.  
For a $\bar{D}^{0}$, we require the reconstructed invariant mass to lie between 1.72 and 2.02 \GeV. For $D^-$ and $D^{*-}$, we require $|M_{D^-} - 1870\ \mathrm{MeV}/c^2| <  10$ \MeV, $|M_{D^{*-}} - 2010\ \mathrm{MeV}/c^2| <  150$ \MeV, and $ |M_{D^{*-}}-M_{\bar{D}^{0}} -145\ \mathrm{MeV}/c^2| < 9$ \MeV, where $M_{D^{(*)-}}$ and $M_{\bar{D}^{0}}$ are the reconstructed masses of $D^{(*)-}$ and $\bar{D}^{0}$, respectively.

We identify the signals using two kinematic variables:  the 
energy difference ($\Delta E$) and the beam-energy-constrained mass ($M_{\rm bc}$),
\begin{equation}
\begin{split}
&\Delta {E}={E}_{B}-{E}_{\rm beam}\\
&{M}_{\rm bc}=\sqrt{{E}_{\rm beam}^{2}-p_{B}^2 c^2}/c^2,
\end{split}
\end{equation}
where $E_{B}$ and $p_{B}$ are the energy and momentum of the $B$ meson and $E_{\mathrm{beam}}$ is the beam energy, all measured in the $\Upsilon(4S)$ center-of-mass (CM) frame.

%\section{Background suppression and signal extraction}\label{sec_bkg_discrimination_signal_extra}
We optimize all selection criteria using Monte Carlo (MC) event samples before examining the data. These samples, both for signal and background, are generated using EvtGen \cite{EvtGen} and later processed with a GEANT3-based detector simulation program 
that provides the detector-level information \cite{Geant}.

Using the generated MC samples, the fit region is defined as $ -0.1\ \mathrm{GeV} < \Delta E < 0.3\ \mathrm{GeV}$ and  $5.22$ GeV/$c^2 <M_{\rm bc} < 5.30$ GeV/$c^2$ while the signal region is defined as $|\Delta E| < 0.05\ \mathrm{GeV}$ and  $5.27$ GeV/$c^2 <M_{\rm bc} < 5.29$ GeV/$c^2$.

Two major sources contribute as background: $e^{+}e^{-} \rightarrow q\bar{q}\ (q=u,\ d,\ s,\ c)$ production, also known as the continuum background, and other $b\rightarrow c$ dominated $B$ meson decays, labeled generically as $B$ decays in this paper. 

To suppress the continuum background, we use the difference between its jet-like topology and the spherical $B$-decay topology. We calculate the distributions of 23 modified Fox-Wolfram moments from the final-state particle momenta given by the signal and background MC \cite{FoxWolfarmMoment,ksfw}. A Fisher discriminant that enhances the signal and background separation with a weighted linear combination of the moments is then calculated \cite{fisher}. We augment the obtained probability density functions (PDFs) of the Fisher discriminant for the signal and background with two more variables to form the signal (background) likelihood ${\mathcal L}_{\mathrm{S(B)}}$: the axial distance ($\Delta z$)
 between the vertices of the candidate $B$ and the remaining final-state particles --- presumably from the other $B$ --- and the cosine of the polar angle of the $B$ momentum (${\mathrm{cos}}\theta_{B}$) in the CM frame. The PDFs used for the modified Fox-Wolfram moments, $\Delta z$, and $\mathrm{cos}\theta_{B}$ are bifurcated Gaussian functions, the sums of three Gaussian functions, and second-order polynomials, respectively.

To suppress the background, we optimize the selection criteria 
for  $[{\mathcal L}_{\mathrm{S}}/({\mathcal L}_{\mathrm{S}}+{\mathcal L}_{\mathrm{B}})]_{D(D^{*})}<\alpha_{D(D^{*})}$, $|{M}_{D^{-}} - 1870\ \mathrm{MeV}/c^{2}|<\beta_{D}$ \MeV, 
and $|{M}_{D^{*-}}-{M}_{\bar{D}^{0}} - 145\ \mathrm{MeV}/c^{2}|<\beta_{D^{*}}$ \MeV~simultaneously and obtain $\alpha_{D}=0.53$, $\alpha_{D^{*}}=0.40$, $\beta_{D}=10$, and $\beta_{D^{*}}=9$. The $\beta$ selections correspond to $\pm2.4\sigma$ and $\pm12.4\sigma$ selections around the nominal $M_{D^{*-}}$ and $M_{D^{*-}}-M_{\bar{D}^0}$. This procedure maximizes the figure of merit, $N_{\mathrm{S}}/\sqrt{N_{\mathrm{S}}+N_{\mathrm{B}}}$,
where $N_{\mathrm{S}}$ and $N_{\mathrm{B}}$ are the expected yields of signal and background, respectively, in the signal region. We use the theoretical expectations in Eq.~(1) to obtain $N_{\mathrm{S}}$ and normalize the $q\bar{q}$
and generic $B$ MC samples to the integrated luminosity to obtain $N_{\mathrm{B}}$. After applying all the selection criteria, the fractions of events with multiple signal candidates are found to be 3.5\% and 5.6\% in the $D$ and $D^*$ modes, respectively. To ensure that no event has multiple entries in the fit region, we retain the $B$ candidate with the smallest vertex-fit $\chi^2$ in each event, where the vertex-fit is performed using all charged tracks from the $B$ candidate except those from $\bar{\Lambda}$.

We model the signal $\Delta E$ distribution with the sum of three Gaussian functions; and the ${M}_{\rm bc}$ distribution with the sum of two Gaussian functions. We model the background $\Delta E$ shape with a second-order polynomial; and the ${M}_{\rm bc}$ shape with an ARGUS function \cite{Albrecht90}. We determine the PDF shapes with MC samples and calibrate the means and widths of the signal PDFs using a large control sample of $B^{0} \rightarrow \pi^{+} K^{0}_{\mathrm{S}} D^{(*)-}$ decays from the data. The signal yields are extracted separately
from eight di-baryon ($p\bar{\Lambda}$) invariant mass bins, in the ranges of 2.05--3.41 \GeV~for the $D$ mode and 2.05--3.30 \GeV~for the $D^*$ mode. We obtain the signal using a two-dimensional extended unbinned maximum likelihood fit in $\Delta E$ and ${M}_{\rm bc}$.

%\section{results}\label{sec_physics_result}

\begin{figure}[htb!]
\centering
\includegraphics[width=0.6\linewidth]{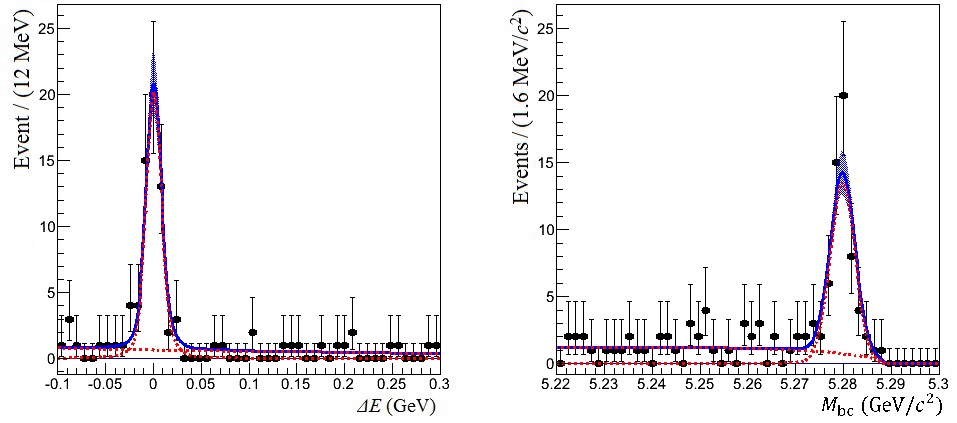}\\
\includegraphics[width=0.6\linewidth]{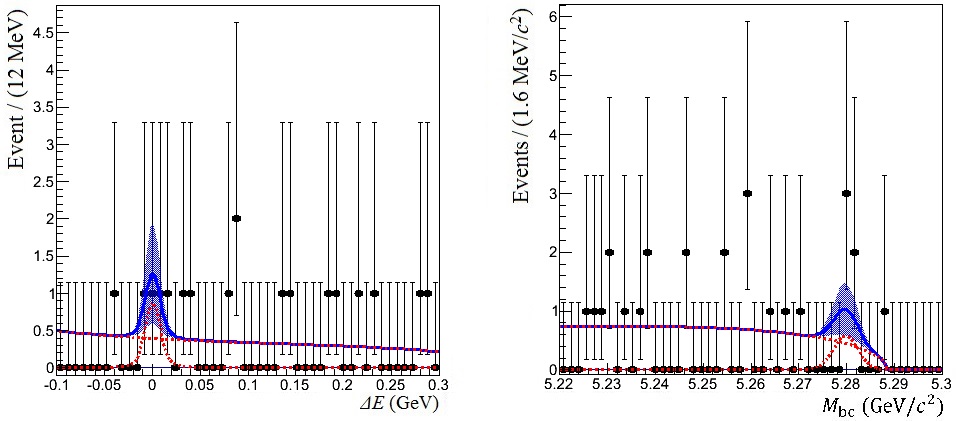}\\
\includegraphics[width=0.6\linewidth]{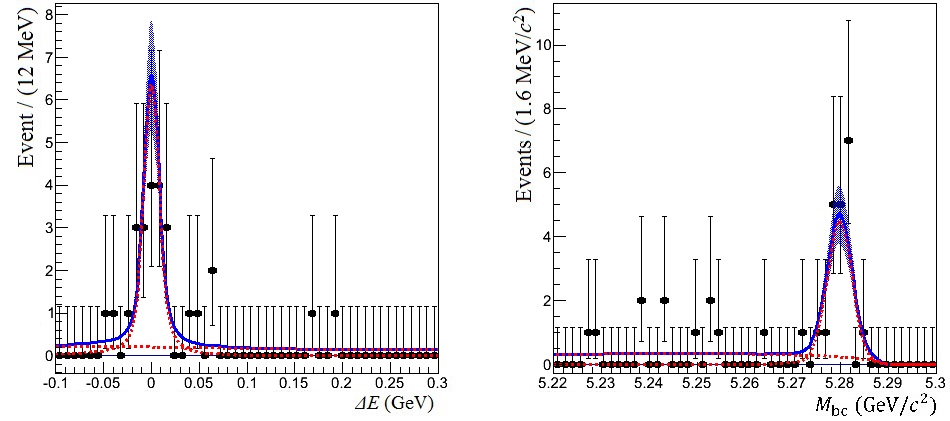}\\
\includegraphics[width=0.6\linewidth]{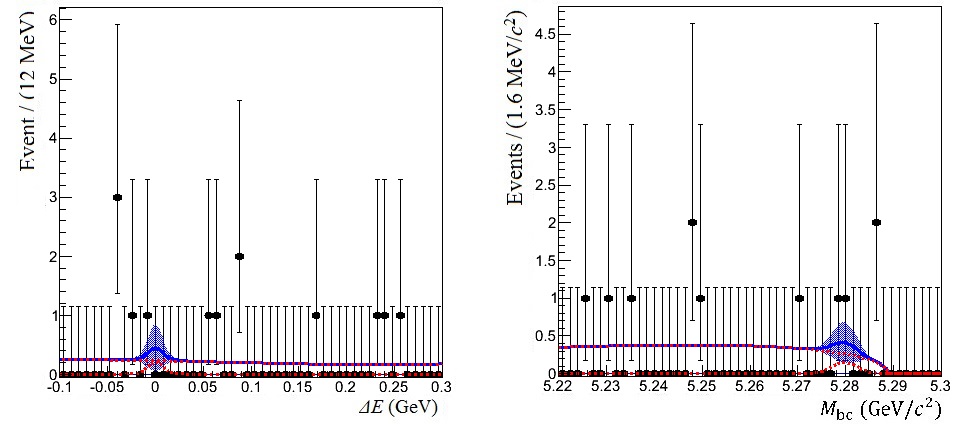}
%\captionsetup{justification=justified}%width=1.0\linewidth}
\caption{Projections of typical $\Delta E$-$M_{\rm bc}$ fits to data for events in the signal region of the orthogonal variable. The peaking and flat red dotted lines represent the signal and background components; the blue solid lines with the dotted areas represent the combined PDFs with their $1\sigma$ uncertainty bands. The top (bottom) four panels from top to bottom show the fits in the lowest and highest $M_{p\bar{\Lambda}}$ bin in the $D$ ($D^*$) mode.}
\label{binned_fit_example}
\end{figure}

Figure~\ref{binned_fit_example} illustrates the fit results of 
the lowest and highest $p{\bar\Lambda}$ mass bins for 
the $D$ and $D^*$ modes. We observe clear signal peaks with very low background in the lowest $M_{p\bar\Lambda}$ bin, indicating an enhancement near threshold. As the efficiency is dependent on $M_{p\bar\Lambda}$, Table~\ref{binned_yield_table} lists
the efficiencies and fitted yields in all mass bins for the two modes. Note that the efficiencies shown do not include the sub-decay branching fractions.   

\begin{table}[b!]
\centering
%\captionsetup{width=13cm}
\caption{The fitted signal yield and efficiency in each $M_{p\bar{\Lambda}}$ bin. To obtain a stable fit, we combine the last three bins in the $D^*$ mode into the sixth bin.}
\label{binned_yield_table}
\begin{tabular}{c|cc|c|cc}
\hline
\hline
%\multirow{2}{*}{$M_{p\bar{\Lambda}}$ (\GeV)} & \multicolumn{2}{c|}{$D$ mode} & \multirow{2}{*}{$M_{p\bar{\Lambda}}$ (\GeV)} & \multicolumn{2}{c}{$D^*$ mode}\\
%\cline{2-3}\cline{5-6} 
% & Yield & Eff.(\%) & &  Yield & Eff.(\%)\\
$M_{p\bar{\Lambda}}$ & \multicolumn{2}{c|}{$D$ mode} & $M_{p\bar{\Lambda}}$ & \multicolumn{2}{c}{$D^{*}$ mode} \\
\cline{2-3}\cline{5-6}
(\GeV) & Yield & Eff.(\%) & (\GeV) &  Yield & Eff.(\%)\\
\hline
2.05--2.22 & $57\pm8$ & $12.2\pm0.0$ & 2.05--2.21 & $19\pm5$ & $12.2\pm0.0$\\ 
2.22--2.39 & $24\pm5$ & $10.5\pm0.0$ & 2.21--2.36 & $9\pm3$  & $10.2\pm0.0$\\
2.39--2.56 & $14\pm4$ & $9.5\pm0.1$ & 2.36--2.52 & $5\pm3$ & $8.7\pm0.0$\\
2.56--2.73 & $8\pm3$ & $9.8\pm0.1$ & 2.52--2.68 & $2\pm1$ & $8.4\pm0.1$\\
2.73--2.90 & $3\pm2$ & $10.4\pm0.1$ & 2.68--2.83 & $3\pm2$ & $7.6\pm0.1$\\
2.90--3.07 & $7\pm3$ & $10.9\pm0.2$ & 2.83--3.30 & $1\pm1$& $6.3\pm0.1$\\
3.07--3.24 & $1\pm2$ & $10.8\pm0.3$ &  & &\\
3.24--3.41 & $2\pm2$ & $11.4\pm0.7$ &  & & \\
\hline 
Total & $117\pm12$ & & & $39\pm7$ &\\
\hline
\hline
\end{tabular}
\end{table}

\begin{figure}[htb!]
%\begin{minipage}{0.49\linewidth}
	\includegraphics[width=4.2cm]{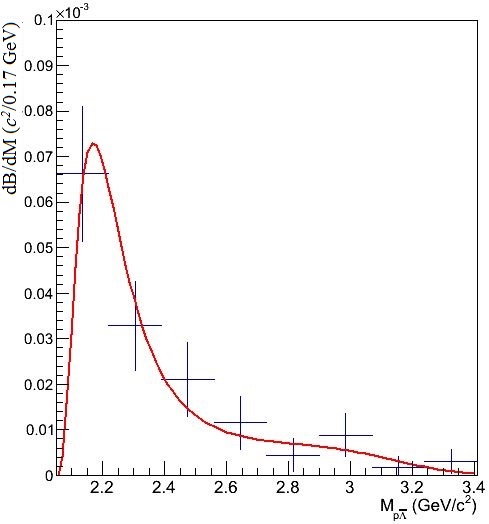}
%\end{minipage}
%\begin{minipage}{0.49\linewidth}
	\includegraphics[width=4.5cm]{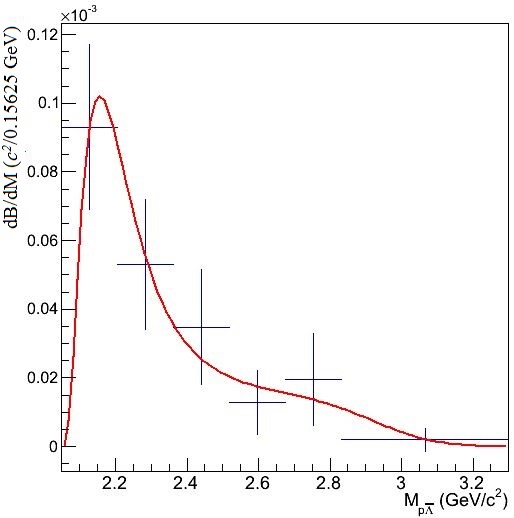}
%\end{minipage}
%\captionsetup{width=12cm}
	\caption{Differential branching fractions of the $D$ (left) and $D^*$ (right) modes in $M_{p\bar{\Lambda}}$. Fit curves are
based on an empirical threshold function (see text).}
\label{differential_efficiency_mLambp}
\end{figure}

Assuming that the branching fractions of $\Upsilon(4S)$ decaying to the charged and neutral $B\bar{B}$ pairs are equal, we use the efficiency and fitted yield in each mass bin to calculate the differential branching fraction and integrate over the entire mass range 
to obtain the branching fraction  $\mathcal{B}=(\sum_{i}N_{i}/ \epsilon_{i})/(\prod\mathcal{B}_{\mathrm{subdecay}} \times N_{B\bar{B}} \times C_{\mathrm{PID}})$,
where $i$ is the mass bin number, $N_{i}$ and $\epsilon_{i}$ are the bin-dependent fitted yield and selection efficiency, respectively, $\mathcal{B}_{\mathrm{subdecay}}$ and  $N_{B\bar{B}}$ are the sub-decay branching fraction and the number of $B\bar{B}$ pairs,
respectively, and $C_\mathrm{PID}$ is the charged-particle identification efficiency correction between MC and data ($0.92$ for the $D$ mode and $0.85$ for the $D^*$ mode). 
Figure\ \ref{differential_efficiency_mLambp} shows the results, where both modes have visible peaks near threshold.
The data are fit with an empirical threshold yield, $m^{a}\times e^{(bm+cm^{2}+dm^{3})}$, \textit{vs.} the mass excess $m=M_{p\bar{\Lambda}}-M_{\bar{\Lambda}}-M_{p}$ by varying $a$, $b$, $c$, and $d$. 
%Table~\ref{binned_yield_table} lists the fitted signal yields and efficiencies in the 
%$M_{p\bar{\Lambda}}$ bins.
The obtained branching fractions are:
\begin{equation}
%\begin{array}{ll}
\begin{split}
&\mathcal{B}(B^{0}\rightarrow p \bar{\Lambda} D^{-})=(25.1\pm2.6\pm3.5)\times10^{-6},\ 19.8\sigma,\\
&\mathcal{B}(B^{0}\rightarrow p \bar{\Lambda} D^{*-})=(33.6\pm6.3\pm4.4)\times10^{-6},\ 10.8\sigma,\\
\end{split}
%\end{array}
\end{equation}
where the quoted uncertainties are statistical and systematic (described later), respectively,
%The signal significance is estimated by evaluating  $\sqrt{
%\sum_i{-2\ln (L_\mathrm{0}/L_\mathrm{max})}}$ with 8 and 6 degrees of freedom representing the number
%of bins in the two modes respectively and $L_\mathrm{max}$ and $L_\mathrm{0}$ being the
%likelihood values with and without the signal component in the fit.
and the significance is estimated by the Z-score of the p-value for $\chi^2=2\sum_i\ln (L_{\mathrm{max},i}/L_{\mathrm{0},i})$ with 8 or 6 degrees of freedom representing the number of bins. $L_\mathrm{max}$ and $L_\mathrm{0}$ are the likelihood values with and without the signal component in the fit, respectively, and $i$ is again the mass bin index. 
The measured branching fractions are clearly incompatible with the theoretical predictions for both the $D$ and $D^*$ modes \cite{C.Chen08}. This indicates that the model parameters used in the calculation need to be revised and, perhaps, some modification of the theoretical framework is required.

\begin{figure}[htb!]
%\begin{minipage}{0.49\linewidth}
	\includegraphics[width=4.3cm,height=4.3cm]{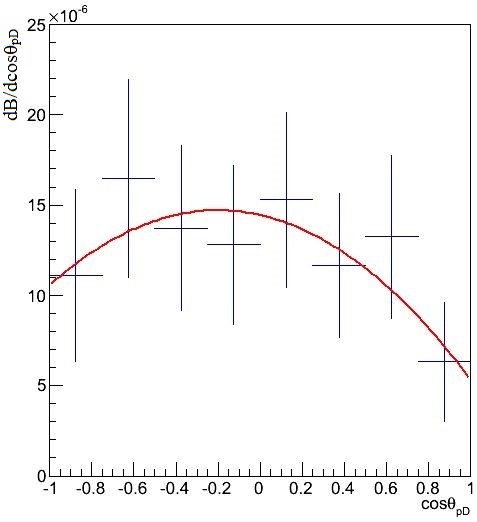}
%\end{minipage}
%\begin{minipage}{0.49\linewidth}
	\includegraphics[width=4.2cm]{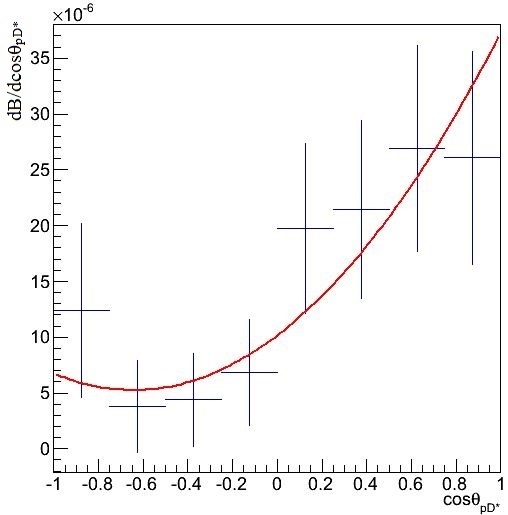}
%\end{minipage}
%\captionsetup{width=12cm}
\caption{Differential branching fractions of the $D$ (left) and $D^*$ (right) modes in $\mathrm{cos}\theta_{pD^{(*)}}$. The fit curves are 
second-order polynomials, as suggested by Ref. \cite{Geng06}.}
\label{differential_efficiency_thetaPD}
\end{figure}

To extract the decay angular distributions, we divide $\mathrm{cos}\theta_{pD^{(*)}}$ into eight bins, where $\theta_{pD^{(*)}}$ is defined as the angle between the proton and meson directions in the $p\bar\Lambda$ rest frame. We follow the same procedure to determine the differential branching fractions in $\mathrm{cos}\theta_{pD^{(*)}}$ as in determining those in $M_{p\bar{\Lambda}}$. 
Table~\ref{binned_theta_yield} lists the fitted signal yields and efficiencies in the $\mathrm{cos}\theta_{pD^{(*)}}$
bins; Fig.\ \ref{differential_efficiency_thetaPD} shows the differential branching fractions. The efficiency is determined with the MC sample, including the threshold enhancement effect as observed in the data.  
\begin{table}[b!]
\centering
%\captionsetup{width=10cm}
\caption{The fitted signal yield and efficiency in each $\mathrm{cos}\theta_{pD^{(*)}}$ bin.}
\label{binned_theta_yield}
\begin{tabular}{c|cc|cc}
\hline
\hline
\multirow{2}{*}{$\mathrm{cos}\theta_{pD^{(*)}}$} & \multicolumn{2}{c|}{$D$ mode} & \multicolumn{2}{c}{$D^*$ mode}\\
\cline{2-5} 
 & Yield & Eff.(\%) & Yield & Eff.(\%)\\
\hline
$-1.00\mbox{\ --\ }-0.75$ & $10\pm4$ & 9.0 & $3\pm2$ & 8.6\\ 
$-0.75\mbox{\ --\ }-0.50$ & $17\pm5$ & 10.5 & $1\pm1$ & 10.2\\
$-0.50\mbox{\ --\ }-0.25$ & $16\pm4$ & 11.5 & $1\pm1$ & 11.3\\
$-0.25\mbox{\ --\ }-0.00$ & $15\pm4$ & 12.2 & $2\pm2$ & 12.2\\
$+0.00\mbox{\ --\ }+0.25$ & $19\pm5$ & 12.8 & $7\pm3$ & 12.7\\
$+0.25\mbox{\ --\ }+0.50$ & $15\pm4$ & 13.0 & $7\pm3$ & 13.0\\
$+0.50\mbox{\ --\ }+0.75$ & $16\pm5$ & 12.6 & $9\pm3$ & 12.8\\
$+0.75\mbox{\ --\ }+1.00$ & $7\pm3$ & 11.5 & $8\pm3$& 11.5\\
\hline
\hline
\end{tabular}
\end{table}
We define the angular asymmetry $A_{\theta}=\frac{\mathcal{B}_{+}-\mathcal{B}_{-}}{\mathcal{B}_{+}+\mathcal{B}_{-}}$, where $\mathcal{B}_{+(-)}$ represents the branching fraction of positive (negative) cosine value. The results are
\begin{equation}
\begin{split}
&A_{\theta}(B^{0}\rightarrow p \bar{\Lambda} D^{-})=-0.08\pm0.10,\\
&A_{\theta}(B^{0}\rightarrow p \bar{\Lambda} D^{*-})=+0.55\pm0.17,\\
\end{split}
\end{equation}
where the uncertainty is purely statistical since the correlated systematic uncertainties cancel in the $A_{\theta}$ calculation. The angular distributions of the $D$ and $D^*$ modes appear to have distinct trends, even though they are both categorized as current-type decays. More data are needed to make the result conclusive.

%\section{Systematic uncertainty}
Three major categories of systematic uncertainties are considered: in the signal yield determination, in the efficiency estimation, and in translating the signal yields and efficiencies into the branching fractions. Table \ref{tab_sys_error} lists all the systematic uncertainties.

We observe a mild peaking background in the $M_{\rm bc}$ fit region due to $B^{+}\rightarrow p \bar{\Lambda} \bar{D}^{*0}$, plausibly by 
the replacement of the low-momentum $\pi^0$ in $\bar{D}^{*0}\rightarrow \bar{D}^{0}\pi^0$ with an unaffiliated $\pi^-$ or $K^-$ to reconstruct a ${D}^{*-}$. To study its contribution to the uncertainty in the $D^{*}$ mode, a dedicated MC sample of this background mode is generated. Based on its current branching fraction upper limit \cite{Chen11}, we subtract 0.5 events from the extracted signal yield and assign $\pm0.5$ events as the systematic uncertainty. We have verified that our signal extraction method is robust and see negligible systematic bias in the signal yield when assuming 0.1 to 10 times the theoretical branching fractions (about 1.6 to 160 events) in an MC ensemble test.

\begin{table}[b!]
\centering
%\captionsetup{width=13cm}
\caption{The systematic uncertainties in the $D$ and $D^*$ modes. The $\approx$ signs indicate the $M_{p\bar{\Lambda}}$ dependence of the uncertainty.}
\label{tab_sys_error}
		\begin{tabular}{c|cc}
\hline
\hline
\multirow{2}{*}{Item} & \multicolumn{2}{c}{Systematic\ uncertainty (\%)}\\
\cline{2-3}
 & $D$ mode & $D^{*}$ mode\\
\hline
Yield bias & negligible & $1.3\ (0.5$ evt.)\\
%\hline
Modeling & $\approx3$ & $\approx2$\\
Charged track & $2.1$ & $4.3$\\
Charged hadron identification & $1.3$ & $1.8$\\
$\bar{\Lambda}$ identification & $4.0$ & $4.4$\\
$M_{D^{-}}$, $M_{D^{*-}}-$$M_{\bar{D}^{0}}$ window & $2.0$ & negligible\\	
${\mathcal L}_{\mathrm{S}}/({\mathcal L}_{\mathrm{S}}+{\mathcal L}_{\mathrm{B}})$ requirement & $11.5$ & $11.0$\\
PDF shape & negligible & negligible\\
N$_{B\bar{B}}$ & $1.4$ & $1.4$\\
Sub-decay $\mathcal{B}$ & $2.2$ & $1.7$\\
\hline
Overall & $13.9$ & $13.1$\\
\hline
\hline
\end{tabular}
\end{table}

For the reconstruction efficiency, we consider the following systematic uncertainties: the signal MC modeling for the threshold enhancement effect using the bound state assumption, charged track reconstruction, charged hadron identification, $\bar{\Lambda}$ reconstruction, background discrimination selections, and the PDF shapes. The modeling uncertainty is estimated by comparing the efficiency calculation based on two different MC samples, one generated assuming $p$-$\bar{\Lambda}$ bound states and the other with three-body phase-space decays, in each $M_{p\bar{\Lambda}}$ bin. As the result is highly threshold-enhanced, we use the efficiency given by the bound-state model to calculate the branching fractions and take the differences as the systematic uncertainties between the two models. The uncertainty is about 3\ (2)\% in the $D\ (D^*)$ mode, depending on the bins. For each charged track except the low-momentum pion in $D^{*-}\rightarrow \bar{D}^{0}\pi^{-}$, a 0.35\% uncertainty is assigned to take into account the data-MC difference in the charged track reconstruction. For the low-momentum pion, a 2.5\% uncertainty is assigned. We use the $\Lambda \rightarrow p \pi^{-}$ and $D^{*+} \rightarrow D^{0}\pi^{+}$, $D^{0} \rightarrow K^{-}\pi^{+}$ 
samples to calibrate the MC \rlap{$\,p$}{${\mathstrut}^{\scriptscriptstyle(-)}$}, $K^{\pm}$, $\pi^{\pm}$ identification efficiencies and assign uncertainties. For the $\bar{\Lambda}$ reconstruction, we estimate the uncertainty by considering the data--MC difference of tracks displaced from the IP, the $\bar{\Lambda}$ proper time, and $\bar{\Lambda}$ mass distributions. The uncertainties due to the $\alpha_{D^{(*)}}$ selections are estimated separately with the control sample mode, $B^{0}\rightarrow \pi^{+}K_{S}^{0}D^{(*)-}$. 
We compare the data--MC efficiency differences with or without 
the $\alpha$ selections, where the non-negligible statistical uncertainties are also included. In both cases, the obtained $\mathcal{B}(B^{0}\rightarrow \pi^{+}K_{S}^{0}D^{(*)-})$ 
is found to be consistent with the world average, indicating overall reliability of our methodology. For the $\beta_{D}$ and $\beta_{D^*}$ selections, 
we compare the widths of the peaking components in $M_{D^{-}}$ and $M_{D^{*-}}- M_{\bar{D}^{0}}$ in the MC and data and quote the differences as the uncertainties. We also relax the shape variables of the signal PDF when fitting the control sample and compare the difference to MC-determined PDF. The resulting difference in the calculated $\mathcal{B}(B^{0}\rightarrow \pi^{+}K_{S}^{0}D^{(*)-})$ is negligible.

In the translation from signal yields to branching fractions, we consider the uncertainties of $\mathcal{B}_{\mathrm{subdecay}}$ and $N_{B\bar{B}}$. The uncertainties of $\mathcal{B}_{\mathrm{subdecay}}$ are obtained from Ref.\ \cite{Olive14}. For $N_{B\bar{B}}$, on- and off- resonance di-lepton events, $e^{+}e^{-}\rightarrow q\bar{q}$ MC and data difference, primary vertex sideband data, and statistical uncertainty are combined to estimate the uncertainty.

%\section{Conclusion}\label{sec_conclusion}
In this paper, we have reported the first observation of the $B^{0}\rightarrow p \bar{\Lambda} D^{-}$ and $B^{0}\rightarrow p \bar{\Lambda} D^{*-}$ decays with branching fractions $(25.1\pm2.6\pm3.5)\times10^{-6}\ (19.8\sigma)$ 
and $(33.6\pm6.3\pm4.4)\times10^{-6}\ (10.8\sigma)$. The threshold enhancement effect observed in $M_{p\bar{\Lambda}}$ is found to be consistent with many other three-body baryonic $B$ decays. 
%The obtained branching fractions disagree with predictions based on the factorization approach, as does the ratio of the branching fraction between the $D$ and $D^*$ modes, and also those between the charged and neutral $B$ modes. 
The obtained branching fractions disagree with predictions based on the factorization approach, as do the measured 
ratios of branching fractions, both for the $D$ and $D^*$ modes and for the charged and neutral $B$ modes.
We also find potential angular asymmetry in the $D^*$ mode but not in the $D$ mode. Theoretical explanations, as well as 
confirmation from experiments with sizable data sets, such as LHCb and Belle II, will be needed in the future.

%----------- Long version, for most papers ----------- 
We thank the KEKB group for the excellent operation of the
accelerator; the KEK cryogenics group for the efficient
operation of the solenoid; and the KEK computer group,
the National Institute of Informatics, and the 
PNNL/EMSL computing group for valuable computing
and SINET4 network support.  We acknowledge support from
the Ministry of Education, Culture, Sports, Science, and
Technology (MEXT) of Japan, the Japan Society for the 
Promotion of Science (JSPS), and the Tau-Lepton Physics 
Research Center of Nagoya University; 
the Australian Research Council and the Australian 
Department of Industry, Innovation, Science and Research;
Austrian Science Fund under Grant No.~P 22742-N16 and P 26794-N20;
the National Natural Science Foundation of China under Contracts 
No.~10575109, No.~10775142, No.~10875115, No.~11175187, and  No.~11475187; 
the Ministry of Education, Youth and Sports of the Czech
Republic under Contract No.~LG14034;
the Carl Zeiss Foundation, the Deutsche Forschungsgemeinschaft
and the VolkswagenStiftung;
the Department of Science and Technology of India; 
the Istituto Nazionale di Fisica Nucleare of Italy; 
National Research Foundation (NRF) of Korea Grants
No.~2011-0029457, No.~2012-0008143, No.~2012R1A1A2008330, 
No.~2013R1A1A3007772, No.~2014R1A2A2A01005286, No.~2014R1A2A2A01002734, 
No.~2014R1A1A2006456;
the Basic Research Lab program under NRF Grant No.~KRF-2011-0020333, 
No.~KRF-2011-0021196, Center for Korean J-PARC Users, No.~NRF-2013K1A3A7A06056592; 
the Brain Korea 21-Plus program and the Global Science Experimental Data 
Hub Center of the Korea Institute of Science and Technology Information;
the Polish Ministry of Science and Higher Education and 
the National Science Center;
the Ministry of Education and Science of the Russian Federation and
the Russian Foundation for Basic Research;
the Slovenian Research Agency;
the Basque Foundation for Science (IKERBASQUE) and 
the Euskal Herriko Unibertsitatea (UPV/EHU) under program UFI 11/55 (Spain);
the Swiss National Science Foundation; the Ministry of Science and Technology
and the Ministry of Education of Taiwan; and the U.S.\
Department of Energy and the National Science Foundation.
This work is supported by a Grant-in-Aid from MEXT for 
Science Research in a Priority Area (``New Development of 
Flavor Physics'') and from JSPS for Creative Scientific 
Research (``Evolution of Tau-lepton Physics'').

\bibliographystyle{apsrev4-1}
	\bibliography{paper_citation}

\end{document}